\documentclass[doublecol]{epl2}

\title{On statistics of molecular chaos}
\shorttitle{On statistics of molecular chaos}

\author{Yuriy E. Kuzovlev}
\shortauthor{Yu. E. Kuzovlev}

\institute{Donetsk Institute for Physics and Technology of NASU -
ul.\,R.\,Luxemburg 72, Donetsk 83114, Ukraine}


\abstract{It is shown that the BBGKY equations for a particle
interacting with ideal gas imply exact relations between probability
distribution of path of the particle, its derivatives in respect to
the gas density and irreducible many-particle correlations of gas
atoms with the path. These relations visualize that the correlations
of any order always significantly contribute to evolution of the path
distribution, so that the exact statistical mechanics theory does not
reduce to the classical kinetics even in the low-density (or
Boltzmann-Grad) limit.}

\begin{document}

\maketitle

\section{\,Introduction}

In the seminal book \cite{bog}  Bogoliubov formulated powerful tools
for statistical theory of transport processes and noises in
many-particle systems, in particular, the exact hierarchy of
evolution equations (now referred to as the
Bogolyubov-Born-Green-Kirkwood-Yvon, or BBGKY, hierarchy \cite{re})
for $\,s\,$-particle ($\,s=1,2,\dots\,$) distribution functions. But
in practice the theory was canceled out by common prejudice that it
must confirm, at least for dilute gases, the Boltzmann's ``molecular
chaos hypothesis'', - the ``Sto{\ss}zahlansatz'' \cite{bol}, - and
thus reduce to the Boltzmann equation. This idea is so much
attractive that Bogoliubov introduced own hypotheses which allow to
truncate the BBGKY hierarchy at $\,s=2\,$ and perform the desired
reduction \cite{bog}. In fact, however, until now neither Boltzmann's
nor Bogoliubov's assumptions have a rigorous substantiation based on
the BBGKY hierarchy itself\,\footnote{\, We take in mind spatially
inhomogeneous gas evolutions, of course, since in respect to strictly
homogeneous (translation invariant) dilute gas the Sto{\ss}zahlansatz
is undoubtedly true.}\,, and hence the Boltzmann equation also stays
without substantiation\,\footnote{\, The frequently mentioned Lanford
theorem \cite{lan} about gas of hard spheres under the Boltzmann-Grad
limit concerns the so called ``hard-sphere BBGKY hierarchy'' (about
it see also e.g. \cite{re,vblls}) which is not a true BBGKY hierarchy
since represents interactions of the spheres by invented terms like
the Boltzmann collision integrals (i.e. postulates what should be
proved, if any).

We here consider regular smooth interaction potentials only and thus
the true BBGKY equations where the interaction is represented by the
forces (potential gradients) like in the parent Liouville equation.
But our conclusions naturally extend to the limit of singular
``hard-sphere'' interaction. }\,.

Here, I suggest a short visual proof of that all this never will be
substantiated, even for arbitrary dilute gas (or Boltzmann-Grad gas).
At that, in order to simplify the proof and at once strengthen it,
instead of the usual gas I consider a test particle in
thermodynamically equilibrium ideal gas whose atoms interact with
this particle only but not with each other (thus concentrating on
situation least favorable for inter-particle correlations).

Our first step will be representation of the BBGKY hierarchy of
equations for this system in terms of irreducible $\,s$-particle
correlations between the test particle and $\,s-1\,$ gas atoms
($\,s=1\,$ corresponds to probability distribution of variables of
the test particle). Then, we discover that these equations produce a
set of formally exact relations between any $\,s$-particle
correlation and derivative of the previous $\,(s-1)$-particle one in
respect to the gas density. Next, make sure that all the correlations
keeps non-zero even under the low-density (or Boltzmann-Grad) limit
and, consequently, no truncation of the BBGKY hierarchy can be made
without qualitative detriment to its solution. Finally, we touch upon
important properties of the correlations.

\section{\,The BBGKY equations and cumulant distribution functions}
We are interested in random walk $\,{\bf R}(t)\,$ of a test molecule
(TM) which is placed into thermodynamically equilibrium gas and
starts at initial time moment $\,t=0\,$ from a specified position
$\,{\bf R}(0)={\bf R}_0\,$.

Let $\,{\bf P}\,$ and $\,M\,$ denote momentum and mass of TM,
$\,m\,$, $\,{\bf r}_j\,$ and $\,{\bf p}_j\,$ ($\,j=1,2,...\,$) are
masses, coordinates and momenta of gas atoms,\, $\,\Phi({\bf r})\,$
is potential of (short-range repulsive) interaction between any of
them and TM, and $\,n\,$ is gas density (mean concentration of
atoms). At arbitrary time $\,t\geq 0\,$, full statistical description
of this system is presented by the chain of $\,(k+1)$-particle
distribution functions ($\,k=0,1,2,...\,$):\, $\,F_0(t,{\bf R},{\bf
P}|\,{\bf R}_0\,;n\,)\,$\, which is normalized (to unit) density of
probability distribution of TM's variables, and \,$\,F_k(t,{\bf R},
{\bf r}^{(k)},{\bf P},{\bf p}^{(k)}|\,{\bf R}_0\,;n\,)\,$\, (where\,
$\,{\bf r}^{(k)} =\{{\bf r}_1...\,{\bf r}_k\,\}\,$, $\,{\bf p}^{(k)}
=\{{\bf p}_1...\,{\bf p}_k\,\}\,$) which are probability densities of
finding TM at point $\,{\bf R}\,$ with momentum $\,{\bf P}\,$ and
simultaneously finding out some $\,k\,$ atoms at points $\,{\bf
r}_j\,$ with momenta $\,{\bf p}_j\,$. A rigorous definition of such
distribution functions (DF) was done in \cite{bog}. In respect to the
coordinates $\,{\bf r}_j\,$ they are not normalized, but instead (as
in \cite{bog}) satisfy the conditions of vanishing of inter-particle
correlations under spatial separation of particles (in other words,
DF satisfy a cluster property with respect to spacial variables).
Subject to the symmetry of DF in respect to $\,x_j= \{{\bf r}_j,{\bf
p}_j\}\,$\, these conditions can be compactly written as follows:\,
$\,F_k\,\rightarrow\,F_{k-1}\,G_m({\bf p}_k)\,$\, at \,$\,{\bf
r}_k\rightarrow\infty\,$\,,\, where $\,G_m({\bf p})\,$ is the Maxwell
momentum distribution of a particle with mass $\,m\,$.

The enumerated DF obey the standard equations \cite{bog}:
\begin{equation}
\frac {\partial F_k}{\partial t}=[\,H_{k},F_k\,]+n \,\frac {\partial
}{\partial {\bf P}}\int_{k+1}\!\!\Phi^{\,\prime}({\bf R}-{\bf
r}_{k+1}) \,F_{k+1}\,\,\,,\label{fn}
\end{equation}
with $\,k=0,1, ...\,$\, and along with obvious initial conditions
\begin{equation}
\begin{array}{c}
F_k|_{t=\,0}\,=\delta({\bf R}-{\bf R}_0)\, \exp{(-H_k/T\,)}= \label{ic}\\
= \delta({\bf R}-{\bf R}_0)\,G_M({\bf P}) \prod_{j\,=1}^k E({\bf
r}_j-{\bf R})\, G_m({\bf p}_j)\,\,,
\end{array}
\end{equation}
where\, $\,H_{k}\,$ is Hamiltonian of subsystem ``\,$k$ atoms +
TM\,'',\, $\,[...,...]\,$ means the Poisson brackets,\, $\,\int_k ...
=\int\int ...\,\,d{\bf r}_k\,d{\bf p}_k\,$\,,\,
$\,\Phi^{\,\prime}({\bf r}) =\nabla\Phi({\bf r})\,$\,, and $\,E({\bf
r})=\exp{[-\Phi({\bf r})/T\,]}\,$. Notice that one can treat TM as
molecule of non-uniformly distributed impurity and (\ref{fn}) as
equations of two-component gas \cite{bog} in the limit of infinitely
rare impurity with main component being in spatially uniform and
thermodynamically equilibrium state.

Equations (\ref{fn}) together with (\ref{ic}) unambiguously determine
evolution of $\,F_0\,$ and eventually probability distribution of
total TM's displacement, or path, $\,{\bf R}-{\bf R}_0\,$. These
equations will become more transparent if we make a proper linear
change of DF $\,F_k\,$ by new functions $\,V_k\,$\,, namely, with the
help of recurrent relations as follow:
\begin{equation}
\begin{array}{c}
F_0(t,{\bf R},{\bf P}| \,{\bf R}_0;n)\,= \,V_0(t,{\bf R},{\bf P}|\,
{\bf R}_0;n)\,\,\,,\\ F_1(t,{\bf R}, {\bf r}_1,{\bf P},{\bf
p}_1|\,{\bf
R}_0;n)\,=\,\\
=\,V_0(t,{\bf R},{\bf P}|\,{\bf R}_0;n) \,f({\bf r}_1\!-{\bf R},{\bf
p}_1)\,+\\ +\, V_1(t,{\bf R}, {\bf r}_1,{\bf P},{\bf p}_1|\,{\bf
R}_0;n)\,\,\,, \label{cf1}
\end{array}
\end{equation}
where\, $\,f({\bf r},{\bf p}) = E({\bf r})\,G_m({\bf p})\,$\,,
\begin{eqnarray}
F_2(t,{\bf R},{\bf r}^{(2)},{\bf P},{\bf p}^{(2)}|{\bf R}_0;n)\, =
\nonumber\\ =\,V_0(t,{\bf R},{\bf P}|{\bf R}_0;n) \,f(\rho_1,{\bf
p}_1)\,f(\rho_2,{\bf p}_2)\,+\nonumber \\+\,V_1(t,{\bf R},{\bf
r}_1,{\bf P},{\bf p}_1|{\bf R}_0;n) \,f(\rho_2,{\bf p}_2)
+\nonumber\\ + \,V_1(t,{\bf R},{\bf r}_2,{\bf P},{\bf p}_2|{\bf
R}_0;n)
\,f(\rho_1\!,{\bf p}_1)\,+\nonumber \\
+ \,V_2(t,{\bf R},{\bf r}^{(2)}, {\bf P},{\bf p}^{(2)}| \,{\bf
R}_0;n)\,\,\,,\nonumber
\end{eqnarray}
where\, $\,\rho_j\,\equiv\, {\bf r}_j\!-{\bf R}\,$,\, and so on.

Apparently, from viewpoint of the probability theory,\, $\, V_k\,$
are a kind of cumulants, or semi-invariants, or cumulant functions
(CF). It is important to notice that if all these CF were zeros then
all conditional DF of gas, $\,F_k/F_0\,$,\, would be independent on
initial position $\,{\bf R}_0\,$ of TM and thus on its displacement
$\,{\bf R}-{\bf R}_0\,$. This fact makes visible very interesting
speciality of the CF\, $\,V_k\,$\,:\, they are not mere correlations
between instant dynamic states of TM and $\,k\,$ gas atoms but
simultaneously their irreducible correlations with the total past
TM's displacement.

\section{\,Relation of many-particle correlations to probability
law of diffusion of the test particle} In terms of CF the BBGKY
hierarchy acquires tridiagonal structure (we omit uninteresting
algebraic details):
\begin{eqnarray}
\frac {\partial V_{k}}{\partial t}=[H_k,V_k]+n \,\frac {\partial
}{\partial {\bf P}}\int_{k+1}\!\! \Phi^{\,\prime}({\bf
R}-{\bf r}_{k+1})V_{k+1}+\nonumber\\
+\,T\sum_{j\,=1}^{k}\, \mathrm{P}_{kj}\,G_m({\bf p}_k)
\,E^{\prime}({\bf r}_k-{\bf R}) \left[\frac {{\bf P}}{MT}+\frac
{\partial }{\partial {\bf P}}\right ] V_{k-1}\,\,\,.\label{vn}
\end{eqnarray}
Here\, $\,E^{\prime}({\bf r})=\nabla E({\bf r})\,$,\, and\,
$\,\mathrm{P}_{kj}\,$\, symbolizes transposition of the pairs of
arguments $\,x_j\,$\, and $\,x_k\,$. At that, initial conditions
(\ref{ic}) and the above-mentioned clustering conditions \cite{bog}
take extremely simple form:
\begin{equation}
\begin{array}{c}
V_0(0\,,{\bf R},{\bf P}|\,{\bf R}_0;\,n)\,=\delta({\bf R}-{\bf
R}_0) \,G_M({\bf P})\,\,,\\
V_{k}(0\,,{\bf R}, {\bf r}^{(k)},{\bf P},{\bf p}^{(k)}|\,{\bf
R}_0;n)=0\,\,,\label{icv}\\ V_k(t,{\bf R}, {\bf r}^{(k)},{\bf P},{\bf
p}^{(k)}|{\bf R}_0;n)\rightarrow0\,\,\,\,\,\,\texttt{at}\,\,\,\,{\bf
r}_j\rightarrow \infty
\end{array}
\end{equation}
($\,1\leq j\leq k\,$). Thus, as it should be with cumulants,
$\,V_k\,$ ($k>0$) vanish under removal of already one of atoms.

From equations (\ref{vn}) as combined with the boundary and initial
conditions (\ref{icv}) it is clear that passage to the limit in
(\ref{icv}) must realize in an integrable way, so that integrals\,
$\,\widetilde{V}_{k}=\int_{k+1} V_{k+1}\,$\, take finite values. Let
us consider them. By applying the operation $\,\int_k\,$ to equations
(\ref{vn}) one obtains
\begin{eqnarray}
\frac {\partial \widetilde{V}_{k}}{\partial t}=[H_k,\widetilde{V}_k]+n
\,\frac {\partial }{\partial {\bf P}}\int_{k+1}\!\!
\Phi^{\,\prime}({\bf
R}-{\bf r}_{k+1})\,\widetilde{V}_{k+1}\,+\nonumber\\
+\,\,\frac {\partial }{\partial {\bf P}}\int_{k+1}\!\!
\Phi^{\,\prime}({\bf R}-{\bf r}_{k+1})\,V_{k+1}\,+\,\,\,\,\,\,\,\,\,
\,\,\,\,\, \,\,\,\,\,\, \label{vn1}\\
+\,T\sum_{j\,=1}^{k}\, \mathrm{P}_{kj} \,G_m({\bf p}_k) \
\,E^{\prime}({\bf r}_k-{\bf R}) \left[\frac {{\bf P}}{MT}+\frac
{\partial }{\partial {\bf P}}\right ] \widetilde{V}_{k-1}\nonumber
\end{eqnarray}
(with\, $\,k=0,1,...\,$). Because of (\ref{icv}) initial conditions
to these equations are zero:\, $\,\widetilde{V}_{k}(t=0)=0\,$\, at any
$\,k\,$.

Now, in addition to $\,\widetilde{V}_{k}\,$, let us consider
derivatives of CF in respect to the gas density,\, $\,V^{\prime}_{k}=
\partial V_{k}/\partial n\,$\,. It is easy to see that differentiation
of (\ref{vn}) in respect to $\,n\,$ yields equations for the
$\,V^{\prime}_{k}\,$ which exactly coincide with (\ref{vn1}) after
changing there $\,\widetilde{V}_{k}\,$ by $\,V^{\prime}_{k}\,$.
Besides, in view of (\ref{icv}), initial conditions to these
equations again all are equal to zero:\,
$\,V^{\prime}_{k}(t=0)=0\,$\, at any $\,k\geq 0\,$. These
observations strictly imply the equalities\,
$\,V^{\prime}_{k}=\widetilde{V}_{k}\,$,\, or
\begin{eqnarray}
\frac {\partial }{\partial n}\,\, V_{k}(t,{\bf R}, {\bf r}^{(k)},{\bf
P},{\bf p}^{(k)}|\,{\bf R}_0;n)\,=\,\, \,\,\,\, \label{me}\\
=\,\int_{k+1} V_{k+1}(t,{\bf R}, {\bf r}^{(k+1)},{\bf P},{\bf
p}^{(k+1)}|\,{\bf R}_0;n)\,\,\,.\nonumber
\end{eqnarray}
This is main, original and formally exact, result of the present
paper. Evidently, it confirms the assumed finiteness of integrals\,
$\,\widetilde{V}_{k}=\int_{k+1} V_{k+1}\,$\,. Notice also that it
anticipates similar but more complicated general statements of
statistical kinetics of fluids.

\section{\,Discussion and resume}
The equalities (\ref{me}) contain the proof promised in the beginning
of this paper. Indeed, they show, firstly, that all the many-particle
correlations between gas atoms and total path, or displacement, of
the test molecule (TM) really exist, i.e. differ from zero. Secondly,
integral values of all the correlations, represented in the natural
dimensionless form, have roughly one and the same order of magnitude.
Indeed, multiplying equalities (\ref{me}) by $\,n^k\,$ and
integrating them over TM's momentum and all gas variables, we have
\begin{eqnarray}
n^k\, V_k(t,\Delta;n)\,\equiv \,n^k \!\int_1\!\! ...\!\int_k \int\!
V_{k}\,d{\bf P}\,=\nonumber\\ =\, n^k\,\frac {\partial^k
V_0(t,\Delta;n)}{\partial n^k} \,\sim\, c_k
V_0(t,\Delta;n)\,\,,\,\,\nonumber
\end{eqnarray}
where\, $\,V_0(t,\Delta;n)  =\int V_0(t,{\bf R},{\bf P}|{\bf
R}_0;n)\,d{\bf P}\,$\, is just the probability density distribution
of the TM's displacement,\, $\,\Delta ={\bf R}-{\bf R}_0\,$\,,
and\, $\,c_k\,$ some numeric coefficients obviously
comparable with unit. Equivalently, unifying all CF
into one generating function, we can write
\begin{eqnarray}
V_0(t,\Delta;n)+ \sum_{k=1}^{\infty} \frac {u^k n^k}{k!}\,
V_k(t,\Delta;n)\,=\nonumber\\ =\, V_0(t,\Delta;(1+u)\,n)\,\,\,.
\label{gf}
\end{eqnarray}
We see that distribution of total of the TM's random walk ``is made
of its correlations with gas atoms'' like the walk itself is made of
collisions with them. Hence, none of the correlations can be
neglected if we aim at adequate analysis of solution of the BBGKY
equations.

We see also that characteristic spacial volume occupied by the
correlations has an order of the specific volume: $\,(\, |\int_1
...\int_k \int V_{k}\,d{\bf P} \,|/V_0 \,)^{1/k} \,\sim\,
n^{-1}\,$.\, In the Boltzmann-Grad limit, $\,n\rightarrow
\infty\,$,\, $\,r_0\rightarrow 0\,$ ($\,r_0\,$ is effective radius of
the interaction),\, $\,\pi r_0^2 n =1/\lambda =\,$const\,,\, this
volume becomes vanishingly small as measured by the TM's mean free
path $\,\lambda\,$. But, nevertheless, it remains on order of the
effective ``volume of collision'', $\,\sim \pi r_0^2\lambda\,$. This
observation prompts that $\,k$-particle correlations are concentrated
just at those particular subsets of $\,k$-particle phase space which
correspond to collisions. Therefore effects of the correlations
completely hold under the limit. The same is said by the equality
(\ref{gf}) which also holds out. This becomes obvious if we take into
account that actually important parameter of the integrated CF under
the Boltzmann-Grad limit must be $\,\lambda\,$ instead of $\,n\,$ and
rewrite (\ref{gf}) in the form
\begin{eqnarray}
W_0(t,\Delta;\lambda)+
\sum_{k=1}^{\infty} \frac {u^k }{k!}\, W_k(t,\Delta;\lambda)=
W_0(t,\Delta;\lambda/(1+u))\,\,\,, \nonumber
\end{eqnarray}
where\, $\,W_k(t,\Delta;\lambda)=\,\lim\, n^k\,V_k(t,\Delta;n)\,$.
Thus, in essence nothing changes under the  Boltzmann-Grad limit.

It is necessary to underline that the correlations under our
attention are qualitatively different from correlations which appear
in standard approximations of the BBGKY hierarchy and connect
velocities of particles after collision (see e.g. \cite{bal}). In our
notations, a pair correlation of such the kind would look nearly as\,
$\,V_1(t,{\bf R},{\bf r},{\bf P},{\bf p})= F_1(t,{\bf
R}^{\prime},{\bf r}^{\prime}, {\bf P}^{\prime},{\bf p}^{\prime})
-F_1(t,{\bf R},{\bf r},{\bf P},{\bf p})\,$,\, where the primed
variables describe the pre-collision state. Clearly, because of the
phase volume conservation during collision, integration of this
expression over $\,\rho={\bf r}-{\bf R}\,$ and the momenta results in
zero. This observation shows that correlations under our interest do
live not so much in momentum space as in the configuration space.
Their salt is that they connect coordinates and walks of particles
and may coexist with statistical independence of particles'
velocities. By their very nature, they are attributes of spatially
inhomogeneous states and evolutions (evolution of
$\,V_0(t,\Delta;n)\,$ gives an example). More profoundly, statistical
and physical meanings of such correlations were discussed already in
\cite{i1} (so for detail see \cite{i1}\,\footnote{\, Reprint of this
article is available from the arXiv:\, 0907.\,3475\,.}\,).

In view of the aforesaid\,\footnote{\, See also [\,Yu.\,Kuzovlev,\,
Theoretical and Mathematical Physics, 160 (3), 1301 (2009)] and
references therein.} we have to conclude that the Boltzmann-Lorentz
equation \cite{re,vblls} and, moreover, the Boltzmann equation in
itself do not represent a (low-density) limit of the exact
statistical mechanical theory. The classical kinetics is only its
approximate probabilistic model (may be good in one respects but
caricature in others). Of course, in the exact theory also molecular
chaos does prevail. But here it is much more rich, even (and first of
all) in case of dilute gas, and does not keep within naive
probabilistic models.


\end{document}